\documentclass[12pt]{amsart}
\usepackage{amssymb,times}
\usepackage{amsmath}

\numberwithin{equation}{section}

\def\cb{{\mathcal B}}

\def\cs{{\mathcal S}}

\def\bn{{\mathbb N}}

\def\bq{{\mathbb Q}}

\def\bz{{\mathbb Z}}

\def\a{\alpha}
\def\b{\beta}

\def\k{\kappa}
\def\m{\mu}

\def\s{\sigma}

\def\th{\theta}  
 \def\O{\Omega}
\def\z{\zeta}

\newcommand{\be}{\begin{equation}}
\newcommand{\ee}{\end{equation}}
\newcommand{\bes}{\begin{equation*}}
\newcommand{\ees}{\end{equation*}}

\def\h{\mathbf{h}}
\def\zz{\mathbf{s}}

\begin{document}
\title[PHASE TRANSITIONS FOR $P$-ADIC POTTS MODELS]
{ON PHASE TRANSITIONS FOR $P$-ADIC POTTS MODEL WITH COMPETING
INTERACTIONS ON A  CAYLEY TREE}
\author{F.M. Mukhamedov}
\address{F.M. Mukhamedov\\
Departamento de Fisica\\
Universidade de Aveiro\\
Campus Universitário de Santiago\\
3810-193 Aveiro, Portugal} \email{{\tt far75m@yandex.ru,  
farruh@fis.ua.pt}}
\author{U.A. Rozikov}
\address{U.A. Rozikov\\
Institute of Mathematics\\
29,  F.Hodjaev str. \\
Tashkent, 700125, Uzbekistan} \email{{\tt rozikovu@yandex.ru}}
\author{J.F.F. Mendes}
\address{J.F.F. Mendes\\
Departamento de Fisica\\
Universidade de Aveiro\\
Campus Universitário de Santiago\\
3810-193 Aveiro, Portugal} \email{{\tt jfmendes@fis.ua.pt}}

\begin{abstract}
In the paper we considere three state $p$-adic Potts model with
competing interactions on a Cayley tree of order two. We reduce a
problem of describing of the $p$-adic Gibbs measures to the
solution of certain recursive equation, and using it we will prove
that a phase transition occurs if and only if $p=3$ for any value
(non zero) of interactions. As well, we completely solve the
uniqueness problem for the considered model in a $p$-adic context.
Namely, if $p\neq 3$ then there is only a unique Gibbs measure the
model. \vskip 0.3cm \noindent
{\bf Mathematics Subject Classification}: 46S10, 82B26, 12J12.\\
{\bf Key words}: $p$-adic field, Potts model, Cayley tree, Gibbs
measure, phase transition, uniqueness.
\end{abstract}

\maketitle

\section{Introduction}

A non-Kolmogorovian probability models \cite{K2},\cite{K3} in that
probabilities belong to the filed of $p$-adic numbers $\bq_p$ were
developed in connection with $p$-adic quantum models (see for
example, \cite{ADFV},\cite{K1},\cite{MP},\cite{VVZ}). In
\cite{K2},\cite{K4} a measure-theoretical axiomatics of the
$p$-adic probability theory were proceeded. In \cite{K5},\cite{KY}
certain, various limit theorems for $p$-adic valued probabilities
have been proved. In \cite{KL1} the theory of stochastic processes
with values in $p$-adic and more general non-Archimedean fields
having probability distributions with non-Archimedean values has
been developed. There a non-Archimedean analogue of the Kolmogorov
theorem was proved, that gives the opportunity to construct wide
classes of stochastic processes by using finite dimensional
probability distributions. This allowed us to begin  the study and
the development of certain  problems of statistical mechanics in a
context of the $p$-adic probability theory. In
\cite{MR1},\cite{MR2} we have developed the $p$-adic probability
theory approaches to study some models with nearest neighbor
interactions on the Cayley tree, such as Ising and Potts
models\footnote{The classical (real value) contra parts of such
models were considered in \cite{GR}, \cite{KT}}. In those papers
we investigated the set of $p$-adic Gibbs measures and a problem
of phase transitions . Note that the $p$-adic Gibbs measures,
associated with those models, enable Markov property.

In the present paper we consider three state $p$-adic Potts models
with competing interactions on a Cayley tree order two. We note
that the models on a Cayley tree with competing interactions have
been studied extensively (see Refs. \cite{MTA},\cite{SC}) since
the appearance of the Vannimenus model (see Ref.\cite{V}), in
which the physical motivations for the urgency of study such
models was presented. In all of these works no exact solutions of
the phase transition problem were found, but some solutions for
specific parameter values were presented. In this paper we can
completely solve the uniqueness problem for the considered model
in a $p$-adic context. Basically, in the real case there are only
sufficient conditions (like Dobrushin condition \cite{G}) for the
uniqueness of a Gibbs measure of certain models. We show that for
the model under consideration there is a phase transition if and
only if $p=3$ as well.

\section{Preliminaries}

\subsection{$p$-adic numbers and measures}

Let $\bq$ be the field of rational numbers. Throughout the paper
$p$ will be a fixed prime number. Every rational number $x\neq 0$
can be represented in the form $x=p^r\frac{n}{m}$, where
$r,n\in\bz$, $m$ is a positive integer and $p,n,m$ are relatively
prime. The $p$-adic norm of $x$ is given by $|x|_p=p^{-r}$ and
$|0|_p=0$. This norm satisfies so called the strong triangle
inequality
$$
|x+y|_p\leq\max\{|x|_p,|y|_p\}.
$$
This is a ultrametricity of the norm. The completion of $\bq$ with
respect to $p$-adic norm defines the $p$-adic field which is
denoted by $\bq_p$. Let $B(a,r)=\{x\in \bq_p | \ |x-a|_p< r\}$,
where $a\in \bq_p$, $r>0$. By $\log_p$ and $\exp_p$ we mean
$p$-adic logarithm and exponential which are defined as series
with the usual way (see, for more details \cite{Ko}). The domain
of converge for them are $B(1,1)$ and $B(0,p^{-1/(p-1)})$
respectively. These two functions have the following properties
(see  \cite{Ko,VVZ}):
\begin{equation}\label{le}
|\exp_p(x)|_p=1,\ \ \ |\exp_p(x)-1|_p=|x|_p<1, \ \
|\log_p(1+x)|_p=|x|_p<p^{-1/(p-1)} \end{equation} and
\[
 \log_p(\exp_p(x))=x, \ \ \exp_p(\log_p(1+x))=1+x. \]

Let $(X,\cb)$ be a  space, where $\cb$ is an algebra of subsets
$X$. A function $\m:\cb\to \bq_p$ is said to be a {\it $p$-adic
measure} if for any $A_1,...,A_n\subset\cb$ such that $A_i\cap
A_j=\emptyset$ ($i\neq j$)

\[ \mu\bigg(\bigcup_{j=1}^{n} A_j\bigg)=\sum_{j=1}^{n}\mu(A_j).
\]

A $p$-adic measure is called {\it a probability measure} if
$\mu(X)=1$. A $p$-adic probability measure $\m$ is called {\it
bounded} if $\sup\{|\m(A)|_p \ | \ A\in \cb\}<\infty $.

For more detail information about $p$-adic measures we refer to
\cite{K2,K4}.

\subsection{The Cayley tree}

The Cayley tree  $\Gamma^k$ of order $ k\geq 1 $ is an infinite
tree, i.e., a graph without cycles, such that each vertex of which
lies on $ k+1 $ edges . Let $\Gamma^k=(V, \Lambda),$  where $V$ is
the set of vertices of $ \Gamma^k$, $\Lambda$ is the set of edges
of $ \Gamma^k$. The vertices $x$ and $y$ are called {\it nearest
neighbor}, which is denoted by $l=<x,y>$ if there exists an edge
connecting them. A collection of the pairs
$<x,x_1>,...,<x_{d-1},y>$ is called {\it a path} from  $x$ to $y$.
The distance $d(x,y), x,y\in V$  is the length of the shortest
path from $x$ to $y$ in V.

For the fixed $x^0\in V$ we set \[ W_n=\{x\in V | d(x,x^0)=n\}, \
\ V_n=\bigcup_{m=1}^n W_m, \]
\[ L_n=\{l=<x,y>\in L | x,y\in V_n\},\] for a fixed point $ x^0
\in V $.

Denote \[ S(x)=\{y\in W_{n+1} |  d(x,y)=1 \}, \ \ x\in W_n. \] The
defined set is called the set of {\it direct successors}. Observe
that any vertex $x\neq x^0$ has $k$ direct successors and $x^0$
has $k+1$.

Two vertices $x,y\in V$ is called {\it one level
next-nearest-neighboring vertices} if there is a vertex $z\in V$
such that  $x,y\in S(z)$ and they are denoted by $>x,y<$. In this
case the vertices $x,z,y$ are called {\it ternary} and denoted by
$<x,z,y>$.

In the sequel we will consider semi-infinite Cayley tree $J^2$ of
order 2, i.e. an infinite graph without cycles with 3 edges
issuing from each vertex except for $x^0$ and with 2 edges issuing
from the vertex $x^0$.

\subsection{The model}

Let $\bq_p$ be the field of $p$-adic numbers. By $\bq_p^{q-1}$ we
denote $\underbrace{\bq_p\times...\times\bq_p}_{q-1}$. The norm
$\|x\|_p$ of an element $x\in \bq_p^{q-1}$ is defined by
$\|x\|_p=\max\limits_{1\leq i\leq q-1}\{|x_i|_p\}$, here
$x=(x_1,...,x_{q-1})$. By $xy$ we mean the bilinear form on
$\bq_p^{q-1}$ defined by \[ xy=\sum_{i=1}^{q-1}x_iy_i, \ \
x=(x_1,\cdots,x_{q-1}), y=(y_1,\cdots,y_{q-1}).
\]

Let $\Psi=\{\z_1,\z_2,...,\z_q\}$, where $\z_1,\z_2,...,\z_q$ are
elements of $\bq_p^{q-1}$ such that  $\|\z_i\|_p=1$, $i=1,2,...,q$
and
\begin{equation}\label{vec1}
\z_i\z_j= \left\{ \begin{array}{ll}
1, \ \  \textrm{for $i=j$},\\
0, \ \ \textrm{for $i\neq j$}\\
\end{array} \right. (i,j=1,2,...,q-1), \ \z_q=\sum_{i=1}^{q-1}\z_i.
\end{equation}

Let $h\in \bq_p^{q-1}$, then we have $h=\sum_{i=1}^{q-1}h_i\z_i$
and
\begin{equation}\label{vec2}
h\z_i= \left\{\begin{array}{ll}
h_i, \ \ \textrm{for $i=1,2,...,q-1$},\\
\sum_{i=1}^{q-1}h_i, \ \ \textrm{for $i=q$}\\
\end{array} \right.
\end{equation}

We consider the $p$-adic Potts model where spin takes values in
the set $\Psi$ and is assigned to the vertices of the tree
$J^2=(V,\Lambda)$. A configuration $\s$ on $V$ is then defined as
a function $x\in V\to\s(x)\in\Psi$; in a similar fashion one
defines a configuration $\s_n$ and $\s^{(n)}$ on $V_n$ and $W_n$
respectively. The set of all configurations on $V$ (resp. $V_n$,
$W_n$) coincides with $\Omega=\Psi^{V}$ (resp.
$\Omega_{V_n}=\Psi^{V_n},\ \ \Omega_{W_n}=\Psi^{W_n}$). One can
see that $\O_{V_n}=\O_{V_{n-1}}\times\O_{W_n}$. Using this, for
given configurations $\s_{n-1}\in\O_{V_{n-1}}$ and
$\s^{(n)}\in\O_{W_{n}}$ we define their concatenations  by
\[ \s_{n-1}\vee\s^{(n)}=\bigg\{\{\s_n(x),x\in
V_{n-1}\},\{\s^{(n)}(y),y\in W_n\}\bigg\}. \] It is clear that
$\s_{n-1}\vee\s^{(n)}\in \O_{V_n}$.

The Hamiltonian $H_n:\O_{V_n}\to\bq_p$ of the $p$-adic Potts model
with competing interactions has the form
\begin{eqnarray}\label{ham}
H_n(\s)&=&-\sum_{<x,y>\in
L_n}J_{x,y}\delta_{\s(x),\s(y)}-\sum_{>x,y<: x,y\in
V_n}K_{x,y}\delta_{\s(x),\s(y)}\nonumber \\
&&-H\sum_{x\in V_n}\delta_{\z_3,\s(x)}, \ \ n\in\bn,
\end{eqnarray}
here $\s\in\O_{V_n}$, $\delta$ is the Kronecker symbol and
\begin{equation}\label{inter}
\left\{
\begin{array}{lll}
|J_{x,y}|_p<p^{-1/(p-1)}, \ \ \forall <x,y>,\\
|K_{u,v}|_p<p^{-1/(p-1)}, \ \ \forall >u,v<,\\
|H|_p<p^{-1/(p-1)}.
\end{array}
\right.
\end{equation}

\section{Existence of phase transition}

In this section we give a  construction of Gibbs measures for the
three state ($q=3$) $p$-adic  Potts model with competing
interactions on a semi-infinite Cayley tree $J^2$ of order 2, and
establish a phase transition for it.

In the sequel we will assume that the condition \eqref{inter} is
satisfied. Let $\mathbf{h}:x\in V\to h_x\in\bq_p^{q-1}$ be a
function of $x\in V$ such that $\|h_x\|_p<p^{-1/(p-1)}$ for all
$x\in V$. Given $n=1,2,...$ consider a $p$-adic probability
measure $\m^{(n)}_\h$ on $\O_{V_n}$ defined by
\begin{equation}\label{mes}
\mu^{(n)}_\h(\s)=Z^{-1}_{n}\exp_p\{-H_n(\s)+\sum_{x\in
W_n}h_x\s(x)\},
\end{equation}
Here, as before, $\s\in \O_{V_n}$ and $Z_n$ is the corresponding
partition function:
$$
Z_n=\sum_{\tilde\s\in\Omega_{V_n}}\exp_p\{-H(\tilde\s)+\sum_{x\in
W_n}h_x\tilde\s(x)\}.
$$
Note that the measures $\m^{(n)}_\h$ are well defined, since from
\eqref{inter}, $\|h_x\|_p<p^{-1/(p-1)}$ and the strong triangle
inequality one gets
\[
\bigg|H_n(\s)+\sum_{x\in W_n}h_x\s(x)\bigg|_p<p^{-1/(p-1)}
\]
for any $n\in \bn$, which enables the existence of the measures
\eqref{mes}. The compatibility condition for $\m^{(n)}_\h, n\geq
1$ is given by the equality
\begin{equation}\label{comp1}
\sum_{\s^{(n)}\in\O_{W_n}}\m^{(n)}_\h(\s_{n-1}\vee\s^{(n)})=\m^{(n-1)}_\h(\s_{n-1}).
\end{equation}

We note that an analog of the Kolmogorov extension theorem for
distributions can be proved for the $p$-adic measures given by
\eqref{mes} (see \cite{KL1}). Then according to the Kolmogorov
theorem there exists a unique $p$-adic measure $\m_{\mathbf{h}}$
on $\O$ such that for every $n=1,2,...$ and $\s_n\in\O_n$ the
equality holds
\[
\m_\h\bigg(\{\s|_{V_n}=\s_n\}\bigg)=\m^{(n)}_\h(\s_n),
\]
which will be called a {\it $p$-adic Gibbs measure} for the
considered model. It is clear that the measure $\m_\h$ depends on
the function $\h$. By $\cs$ we denote the set of all $p$-adic
Gibbs measures associated with functions $\h=(h_x,\ x\in V)$. If
$|\cs|\geq 2$, then we say that for this model there exists {\it a
phase transition}, otherwise, we say there is {\it no phase
transition} ( here $|A|$ means the cardinality of a set $A$). In
other words,  the phase transition means that there are two
different functions $\h=(h_x,\ x\in V)$ and $\zz=(s_x,\ x\in V)$
for which there exist two $\m_\h$ and $\m_{\zz}$ $p$-adic Gibbs
measures on $\O$, respectively.

Using \eqref{mes} and the argument of the proof of Theorem 3.2
\cite{MR1} we may obtain that the measures $\m^{(n)}_\h, \
n=1,2,...$ satisfy the compatibility condition  \eqref{comp1} if
and only if for any $x\in V$ the following recursive equation
holds:
\begin{equation}\label{comp2}
\left\{
\begin{array}{ll}
h_{x,1}=\log\frac{\th_1F_1\bigg(\th_{xy},\th_{xz},\k_{yz};\exp_p(h_y),\exp_p(h_z)\bigg)}{F_2\bigg(\th_{xy},\th_{xz},\k_{yz};
\exp_p(h_y),
 \exp_p(h_z)\bigg)}\\[6mm]
h_{x,2}=\log
\frac{\th_1F_1\bigg(\th_{xy},\th_{xz},\k_{yz};\exp_p(h_y)^t),
\exp_p((h_z)^t)\bigg)}{F_2\bigg(\th_{xy},\th_{xz},\k_{yz};
\exp_p((h_y)^t), \exp_p((h_z)^t)\bigg)}
\end{array}
\right.
\end{equation}
here $<y,x,z>$ ternary vertices, $\theta_{xy}=\exp_p\{J_{xy}\}$,
$\k_{xy}=\exp_p\{K_{xy}\}$, $\th_1=\exp_p(H)$ and for given vector
$h=(h_1,h_2)$ by $\exp_p(h)$ and $h^t$ we have denoted  the
vectors $(\exp_p(h_1),\exp_p(h_2))$ and $(h_2,h_1)$ respectively,
and $F_i:{\bq_p}^{3}\times {\bq_p}^{4}\to{\bq_p}$, ($i=1,2$)
functions are defined by
\begin{equation}\label{func1}
\left\{
\begin{array}{llll}
F_1(\a_1,\a_2,\b;h,r)&=&\a_1\a_2\b
h_1h_2r_1r_2+\a_1h_1h_2(r_1+r_2)+\a_2r_1r_2(h_1+h_2)\\
&&+\b(h_1r_1+h_2r_2)+h_1r_2+h_2r_1\\[3mm]
F_2(\a_1,\a_2,\b;h,r)&=&\b
h_1h_2r_1r_2+\a_1h_1r_1r_2+\a_2h_1h_2r_1+h_2r_1r_2+h_1h_2r_2\\
&&+\a_1h_1r_2+\a_2h_2r_1+\b h_2r_2+\a_1\a_2\b h_1r_1
\end{array}
\right.
\end{equation}
where $h=(h_1,h_2), r=(r_1,r_2)$.

Consequently, the problem of describing of $\cs$ is reduced to the
finding of solutions of the functional equation \eqref{comp2}.

Write
\[
\Sigma=\{ \h=(h_x\in \bq_p^{2}, x\in V) : h_x \ \
\textrm{satisfies the equation \eqref{comp2}}\}.
\]

To prove the existence of phase transition it suffices to show
that there are two different functions in $\Sigma$. The
description of  arbitrary elements of the set $\Sigma$ is a
complicated problem.

Therefore, assume that $J_{xy}=J$, $K_{xy}=K$ and $H=0$. In this
paper we restrict ourselves to the description of translation -
invariant  elements of $\Sigma$, i.e. in which $h_x=h$ is
independent on $x$.

Let $h_x=h=(h_1,h_{2})$ for all $x\in V$. Then using \eqref{func1}
we can reduce  \eqref{comp2} to the following form
\begin{equation}\label{comp3}
\left\{
\begin{array}{ll}
u_1=\frac{\th^2\k
u_1^2u_2^2+2\th(u_1^2u_2+u_1u_2^2)+\k(u_1^2+u_2^2)+2u_1u_2}{\k
u_1^2u_2^2+2\th u_1^2u_2+2u_1u_2^2+2\th u_1u_2+\k u_2^2+\th^2\k
u_1^2},\\[3mm]
u_2=\frac{\th^2\k
u_1^2u_2^2+2\th(u_1^2u_2+u_1u_2^2)+\k(u_1^2+u_2^2)+2u_1u_2}{\k
u_1^2u_2^2+2\th u_1u_2^2+2u_1^2u_2+2\th u_1u_2+\k u_1^2+\th^2\k
u_2^2},\\[3mm]
\end{array}
\right.
\end{equation}
here $u_1=\exp_p(h_1)$, $u_2=\exp_p(h_2)$ and $\th=\exp_p(J)$,
$\k=\exp_p(K)$.

From \eqref{comp3} it is easily seen that the lines $u_1=u_2$,
$u_1=1$ and $u_2=1$,  are invariant for the equation.  Therefore,
it is enough to consider the equation on the line $u_2=1$, since
other cases can be reduced to this case. So, we rewrite
\eqref{comp3} as follows
\begin{equation}\label{comp4}
u=\frac{(\th^2\k+2\th+\k)u^2+2(\th+1)u+\k}{2(\k+1) u^2+4\th u+
\th^2\k}
\end{equation}

It is evident that  $u=1$ is a solution of \eqref{comp4}, but to
exist a phase transition we are interested for other solutions
one. After some simple algebra we find the following equation
\begin{equation}\label{comp5}
2(\k+1) u^2+(2+2\th-\k-\th^2\k)u-\k=0.
\end{equation}

We have to find a solution of \eqref{comp5} such that
$|u-1|_p<p^{-1/(p-1)}$.

Rewriting \eqref{comp5} as follows
\[
2(\k+1)(u^2-1)+(2+2\th-\k-\th^2\k)(u-1)+2\th-\th^2\k+4=0
\]
we infer that if $|2\th-\th^2\k+4|_p=1$ then \eqref{comp5} does
not have a needed solution.

Therefore, we should require that
$|2\th-\th^2\k+4|\leq\frac{1}{p}$. We know (see \eqref{le}) that
from the properties of the exponential function the parameters
$\th$ and $\k$ satisfy the following inequalities
\begin{equation}\label{th-k}
|\th-1|_p\leq\frac{1}{p}, \ \ \ |\k-1|_p\leq\frac{1}{p}.
\end{equation}

Using these inequalities we derive that the inequality
$|2\th-\th^2\k+4|\leq\frac{1}{p}$ is valid if and only if $p=3$.
Therefore, let us assume that $p=3$. Denote
\[
P(x)=2(\k+1) x^2+(2+2\th-\k-\th^2\k)x-\k.
\]
For $P(x)$ we have $P(1)=2\th-\th^2\k+4$,
$P'(1)=6+2\th-\th^2\k+3\k$, hence by means of \eqref{th-k} one
gets
\[
|P(1)|_3\leq\frac{1}{3}, \ \ \ |P'(1)|_3=1.
\]
So according to the Hensel's lemma (see \cite{Ko}) there is a
solution $U\in\bq_3$ of the equation $P(x)=0$ such that
$|U-1|_3\leq\frac{1}{3}$. Consequently, for the model under
consideration there is a phase transition at $p=3$ for every $J,K$
such that $0<|J|_3\leq\frac{1}{3}$, $0<|K|_3\leq\frac{1}{3}$.

\section{The uniqueness of the Gibbs measure}

From the previous section we infer that if $p\neq 3$ then the
equation \eqref{comp4} has a unique solution. Note that this
solution corresponds to the case $h_x=h$. Therefore, in general,
does there exist a phase transition in this case or not? In this
section we are going to answer this question. Here we will
consider two cases.

\subsection{Non-homogeneous case}

Let us assume that $H=0$ and \eqref{inter} be satisfied. Then it
is not hard to check that $\h=(h_x=0, x\in V)$ is a solution for
\eqref{comp2}.  We will to prove that any other solution of
\eqref{comp2} coincides with this one. To show it we have to
estimate the value $\|h_x\|_p$. Denote $u_{x,i}=\exp_p(h_{x,i})$,
$i=1,2$. Then from \eqref{le} we find
\begin{equation}\label{est}
|h_{x,i}|_p=|u_{x,i}-1|_p, \ \ i=1,2.
\end{equation}
hence we have
\begin{eqnarray}\label{est1}
|u_{x,1}-1|_p&=&\bigg|\frac{F_1(\th_{xy},\th_{xz},\k_{yz};u_y,u_z)-F_2(\th_{xy},\th_{xz},\k_{yz};
u_y, u_z)}{F_2(\th_{xy},\th_{xz},\k_{yz}; u_y, u_z)}\bigg|_p \nonumber\\
&=&\bigg|\k_{yz}u_{y,1}u_{z,1}(\th_{xy}\th_{xz}-1)(u_{y,2}u_{z,2}-1)+u_{y,1}u_{z,2}(\th_{xy}-1)(u_{y,2}-1)\nonumber\\
&&+u_{y,2}u_{z,1}(\th_{xz}-1)(u_{z,1}-1)+u_{y,1}u_{z,1}(\th_{xy}-\th_{xz})(u_{y,2}-u_{z,2})\bigg|_p\nonumber\\
&\leq&\frac{1}{p}\max\bigg\{|u_{y,2}u_{z,2}-1|_p,|u_{y,2}-1|_p,|u_{z,1}-1|_p,|u_{y,2}-u_{z,2}|_p\bigg\}\nonumber\\
&=&\frac{1}{p}\max\bigg\{|h_{y,2}+h_{z,2}|_p,|h_{y,2}|_p,|h_{z,1}|_p,|h_{y,2}-h_{z,2}|_p\bigg\}\nonumber\\
&\leq&\frac{1}{p}\max\bigg\{\|h_{y}\|_p,\|h_{z}\|_p\bigg\}
\end{eqnarray}
here we again used \eqref{le} and
\[
|F_2(\th_{xy},\th_{xz},\k_{yz}; u_y, u_z)|_p=1
\]
which is valid if $p\neq 3$.

Analogously reasoning we derive
\begin{equation}\label{est2}
|u_{x,2}-1|_p\leq\frac{1}{p}\max\bigg\{\|h_{y}\|_p,\|h_{z}\|_p\bigg\}.
\end{equation}

The equality \eqref{est} with \eqref{est1},\eqref{est2} implies
that
\begin{equation}\label{est3}
\|h_x\|_p\leq\frac{1}{p}\max\bigg\{\|h_{y}\|_p,\|h_{z}\|_p\bigg\}.
\end{equation}

Take an arbitrary $\varepsilon>0$. Let $n_0\in\bn$ be such that
$\frac{1}{p^{n_0}}<\varepsilon$. Now iterating \eqref{est3} $n_0$
times one gets
\[
\|h_x\|_p\leq\frac{1}{p^{n_0}}<\varepsilon
\]
this means that $h_x=0$ for all $x\in V$. Thus, the $p$-adic Gibbs
measure is unique.

\subsection{Homogeneous case}

In this subsection we will assume that  $J_{xy}=J$, $K_{xy}=K$ and
$H\neq 0$. Let us suppose that $h_x=h=(h_1,h_2)$ for all $x\in V$.
In this case one easily sees that \eqref{comp2} invariant with
respect to $u_1=u_2$. Therefore, we are looking for a solution of
\eqref{comp2} of the form $(h_1,h_1)$. Then it can be rewritten as
follows

\begin{equation}\label{fix1}
u=\th_1\frac{\th^2\k u^2+4\th u+2(\k+1)}{\k u^2+2(\th+1)u+
\th^2\k+2\th+\k}, \ \ \ u=\exp_p(h_1).
\end{equation}

Denote
\begin{equation}\label{func2}
f(x)=\th_1\frac{\th^2\k x^2+4\th x+2(\k+1)}{\k x^2+2(\th+1)x+
\th^2\k+2\th+\k}.
\end{equation}

Let us show that $f(B(1,p^{-1/(p-1)}))\subset B(1,p^{-1/(p-1)})$.
Indeed, let $|x-1|_p<p^{-1/(p-1)}$, then

\begin{eqnarray*}
|f(x)-1|_p&=&\bigg|\frac{(\th^2\th_1\k-\k)x^2+(4\th\th_1-2\th-2)x+2(\k+1)\th_1-\th^2\k-2\th-\k}{\k
x^2+2(\th+1)x+ \th^2\k+2\th+\k}\bigg|_p\\
&\leq&\max\bigg\{|(\th^2\k\th_1-\k)(x^2-1)|_p,
|(4\th\th_1-2\th-2)(x-1)|_p,\\
&&|(\th^2\k+4\th+2(\k+1))(\th_1-1)|_p\bigg\} \\
&\leq&\max\{|(x-1)|_p,(\th_1-1)|_p\}< p^{-1/(p-1)}
\end{eqnarray*}
here we have used the equality $|\k x^2+2(\th+1)x+
\th^2\k+2\th+\k|_p=1$ which is valid if $p\neq 3$.

Now after some algebra we derive
\begin{eqnarray}\label{fix2}
|f(x)-f(y)|_p&=&\bigg|2\th\k
xy(2-\th-\th^2)\nonumber\\
&&+\k(x+y)(2(\k+1)-\th^2(\th^2\k+2\th+\k))\nonumber\\
&&+4((\th+1)(\k+1)-\th^2\k-2\th-\k)\bigg|_p|x-y|_p\nonumber\\
&\leq&\max\bigg\{|2-\th-\th^2|,\nonumber\\
&&|2(\k+1)-\th^2(\th^2\k+2\th+\k)|_p,\nonumber\\
&&|(\th+1)(\k+1)- \th^2\k-2\th-\k|_p\bigg\}|x-y|_p.
\end{eqnarray}

Using (\ref{th-k}) one can be shown that
\[
\left\{
\begin{array}{lll}
|2-\th-\th^2|\leq\frac{1}{p}, \\
|2(\k+1)-\th^2(\th^2\k+2\th+\k)|_p\leq\frac{1}{p}, \\
|(\th+1)(\k+1)- \th^2\k-2\th-\k|_p\leq\frac{1}{p},
\end{array}
\right.\] The last inequalities with \eqref{fix2} imply that
\begin{equation}\label{fix4}
|f(x)-f(y)|_p\leq\frac{1}{p}|x-y|_p.
\end{equation}

Thus the inequality \eqref{fix4} yields that $f$ is a contraction
of $B(1,p^{-1/(p-1)})$, hence $f$ has a unique fixed point
$\zeta\in B(1,p^{-1/(p-1)})$. Let $\xi=\log_p\zeta$ and
$\bar\xi=(\xi,\xi)$.  Then using the same argument as \eqref{est1}
we may obtain
\[
\|h_{x}-\bar\xi\|_p\leq\frac{1}{p}\max\{\|h_y-\bar\xi\|_p,\|h_z-\bar\xi\|_p\}.
\]

Now using the argument of the previous subsection we get that
$h_x=\bar\xi$ for all $x\in V$.

Thus, the $p$-adic Gibbs measure is unique.

\section{Conclusions}

In the paper we have considered three state $p$-adic Potts model
with competing interactions on a Cayley tree order two. We reduced
a problem of describing of the $p$-adic Gibbs measures to the
solution of certain recursive equation, and using it we proved
that a phase transition occurs if and only if $p=3$ for any value
(non zero) of interactions. If $p\neq 3$ we showed that there is
only a unique Gibbs measure for the inhomogeneous Potts model with
zero external filed, as well as we established that result for
homogeneous model but with non-zero external filed. From these
results we conclude that a phase transition depends only on a
value of $p$. These results totaly different from the real case,
since in this setting there is a phase transition on some
constraints for the interaction parameters (see
\cite{GR},\cite{MR3}). When a $p$-adic Gibbs measure is unique,
then by means of a method of paper \cite{MR1} one can be shown
that the measure is bounded as well.  The results concerning the
uniqueness of the Gibbs measures extend results obtained in
\cite{MR1,MR2}. We hope our results will force to study certain
limit theorems for such kind of measures, since they naturally
appear from some Hamiltonian systems, and on the other hand, these
measures enable a Markov property. We also hope that these
investigations give some opportunity to study Hamiltonian systems
over networks in a $p$-adic setting \cite{DGM}.

\section*{acknowledgments} The authors would like express
their gratitude the organizers of the conference on $p$-adic
Mathematical Physics for an invitation. F.M. thanks the FCT
(Portugal) grant SFRH/BPD/17419/2004. F.M. and U.R thank also for
grant $\Phi$-2.1.56 of CST of Uzbekistan. U.R. acknowledges prof.
M.Cassandro for an invitation to "La Sapienza" University and NATO
Reintegration Grant FEL.RIG.980771.

\end{document}